\documentclass[11pt]{article}
\pdfoutput=1

\usepackage{euscript}
\usepackage{amssymb}
\usepackage{amsfonts}
\usepackage{amsbsy}
\usepackage{epsfig}
\usepackage{amsthm}
\usepackage{amscd}
\usepackage{amstext}
\usepackage{verbatim}
\usepackage{amsmath}
\usepackage{cancel}
\usepackage{capt-of}
\usepackage{empheq}
\usepackage{subfigure}

\usepackage[nosort]{cite}
\usepackage{bm}
\usepackage{authblk}
\usepackage{esint}
\usepackage[T1]{fontenc}
\usepackage{mathdots}
\usepackage[centertableaux]{ytableau}

\usepackage[utf8]{inputenc}
\usepackage{graphicx}
\usepackage{physics}
\usepackage{mathtools}
\usepackage{bbold}

\usepackage{setspace}
\usepackage{colortbl}
\usepackage{xcolor}

\usepackage[pdftex,linktoc=all]{hyperref}
\hypersetup{ 
colorlinks=true, 
linkcolor=black, 
citecolor=red, 
}

\def\ben{\begin{equation}}
\def\een{\end{equation}}
\def\half{{\textstyle{\frac{1}{2}}}}
\let\a=\alpha \let\b=\beta

\let\pa=\partial
\def\be{\begin{equation}}
\def\ee{\end{equation}}
\def\beq{\begin{equation}}
\def\eeq{\end{equation}}
\def\ba{\begin{array}}
\def\ea{\end{array}}

\def\dalemb#1#2{{\vbox{\hrule height .#2pt
       \hbox{\vrule width.#2pt height#1pt \kern#1pt
               \vrule width.#2pt}
       \hrule height.#2pt}}}

\newcommand{\bea}{\begin{eqnarray}}
\newcommand{\eea}{\end{eqnarray}}

\makeatletter
\newcommand*\bigcdot{\mathpalette\bigcdot@{.5}}
\newcommand*\bigcdot@[2]{\mathbin{\vcenter{\hbox{\scalebox{#2}{$\m@th#1\bullet$}}}}}
\makeatother

\renewcommand{\eqref}[1]{(\ref{#1})}

\title{Cosmological quantum states of de Sitter-Schwarzschild \\
are static patch partition functions}
\author{Matthew J. Blacker and Sean A. Hartnoll}
\affil{\it Department of Applied Mathematics and Theoretical Physics, \\
\it University of Cambridge, Cambridge CB3 0WA, UK
}
\date{}

\begin{document}

\maketitle

\begin{abstract}

We solve the Wheeler-DeWitt equation in the `cosmological interior' (the past causal diamond of future infinity) of four dimensional dS-Schwarzschild spacetimes. Within minisuperspace there is a basis of solutions labelled by a constant $c$, conjugate to the mass of the black hole. We propose that these solutions are in correspondence with partition functions of a dual quantum mechanical theory where $c$ plays the role of time. The quantum mechanical theory lives on worldtubes in the `static patch' of dS-Schwarzschild, and the partition function is obtained by evolving the corresponding Wheeler-DeWitt wavefunction through the cosmological horizon, where a metric component $g_{tt}$ changes sign. We establish that the dual theory admits a symmetry algebra given by a central extension of the Poincar\'e algebra $\mathfrak{e}(1,1)$ and that the entropy of the dS black hole is encoded as an averaging of the dual partition function over the background $g_{tt}$.

\end{abstract}

\newpage

\tableofcontents

\section{Introduction}

In this paper we will propose a quantum mechanical framework for the cosmological dynamics of four dimensional gravity with a positive cosmological constant. That is to say, we aim to develop a holographic framework in the absence of a rigid boundary. Rigid boundaries provide a conventional notional of quantum mechanical time evolution \cite{Brown:1992br, Balasubramanian:1999re, Maldacena:2001kr}, but in the absence of a boundary the gravitational Hamiltonian vanishes \cite{DeWitt:1967yk}. This has been an obstacle to the construction of a conventional quantum mechanics for cosmology.

Our starting point will be the observation that cosmological solutions to the semiclassical Wheeler-DeWitt equation involve a constant of integration $c$ that can be interpreted as a time. Wheeler-DeWitt wavepackets that are strongly localized on classical geometries are further characterized by a second constant $M$ that is conjugate to $c$. This second constant is the mass of a Schwarzschild black hole in the de Sitter static patch. We propose that the constant $M$ is the expectation value of the Hamiltonian $H$ of a `dual' conventional quantum mechanics. This Hamiltonian generates translations in $c$. More precisely, as there is no fixed rigid boundary, there is a family of dual quantum mechanical theories for which the Hamiltonian is a function of the induced metric $h$ on the Wheeler-DeWitt slices. The spectra of these Hamiltonians must obey equation (\ref{eq:deltadelta}) below, which states that $H(h)$ has an eigenvalue $M$ whenever $h$ arises as a slice of dS-Schwarzschild with mass $M$.

\begin{figure}[h]
    \centering
    \includegraphics[width=0.75\textwidth]{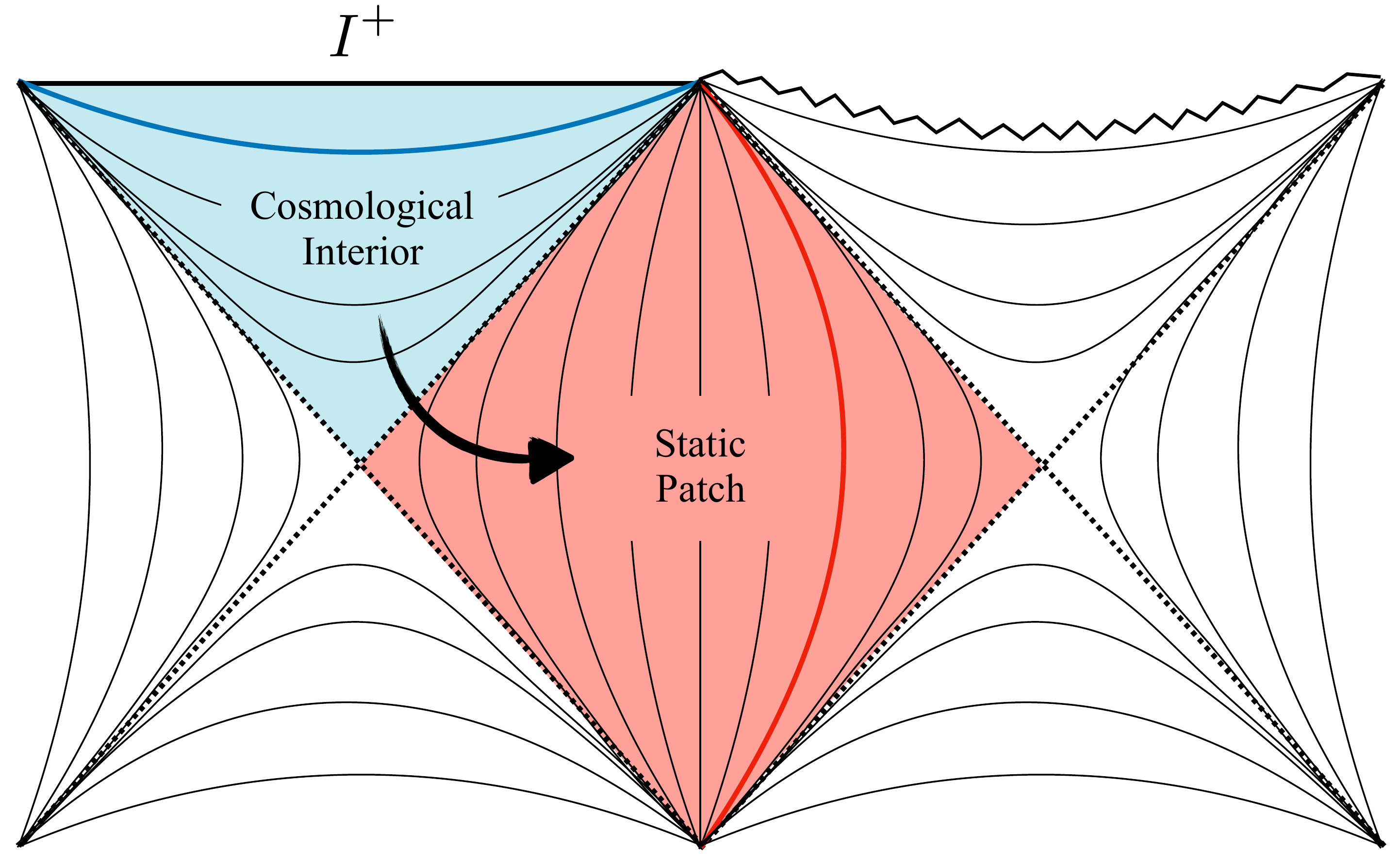}
    \caption{The regions of the Penrose diagram for dS-Schwarzschild that will play a role in our discussion. The Wheeler-DeWitt wavefunction is a function of slices in the (future) cosmological interior. One such slice is shown in dark blue. The continuation of this wavefunction across the cosmological horizon, shown as a diagonal dotted line, will define partition functions in the static patch. These partition functions depend upon worldtube slices in the static patch. One such worldtube is shown in dark red. The static patch is bounded on the other side by the black hole horizon, also shown as a diagonal dotted line.}
    \label{fig:penrose}
\end{figure}

Solutions to the Wheeler-DeWitt equation are usually associated to a causal region of spacetime. We will consider solutions that correspond to the `cosmological interior' of asymptotically de Sitter spacetimes. This is the past causal diamond of future infinity $I^+$, as shown in Fig. \ref{fig:penrose}. The Wheeler-DeWitt wavefunction is a function of the induced Riemmanian 3-metric $h$ on slices of the cosmological interior, $\Psi(h)$. To identify the constant of integration $c$ as a time, we extend the wavefunction $\Psi(h)$ to Lorentzian 3-metrics. This amounts to pulling the wavefunction out through the cosmological horizon and into the static patch, as also shown in Fig. \ref{fig:penrose}. Within the minisuperspace that we will be working, this extension amounts to allowing the metric component $g_{tt}$ of $h$ to be negative. Once in the static patch, we show that solutions to the Wheeler-DeWitt equation can be identified with partition functions of a dual quantum mechanics on the (now Lorentzian) `worldtube' 3-metric $h$, in which $c$ is time.
This correspondence was outlined in the previous paragraph and is stated precisely in equations (\ref{eq:duality}), (\ref{eq:Zgen}), (\ref{eq:path}) and (\ref{eq:full}) below.

The Wheeler-DeWitt states on Lorentzian 3-metrics could also be considered directly as gravitational static patch partition functions, see (\ref{eq:full}), without needing to continue from the cosmological interior. Such a `pure static patch' formulation would be analogous to the AdS/CFT correspondence \cite{Gubser:1998bc, Witten:1998qj}. We have started from the cosmological interior because this is a more familiar setting for the Wheeler-DeWitt equation and because the correspondence then offers a Hamiltonian perspective on the `interior' cosmological quantum states. Every interior state has a corresponding partition function, and thus there is not a unique consistent quantum cosmological state of the universe (cf.~\cite{Hartle:1983ai}).

It is important to emphasize that we are not quantizing the gravitational superspace directly. The partition functions are those of a family of dual theories with Hamiltonians $H(h)$ that live on worldtubes in the static patch. Time evolution generated by $H(h)$ corresponds to evolving between different Wheeler-DeWitt states.

Although we do not identify the dual quantum mechanics explicitly, and hence we are not providing a concrete duality, we are able establish a few properties that the dual theory must have. In section \ref{sec:algebra} we show that dual theory has symmetries given by a central extension of the Poincar\'e algebra $\mathfrak{e}(1,1)$. The coefficient (\ref{eq:coefficient}) of the central extension depends on the radius $R$ of the worldtube. The algebra (\ref{eq:algebra}) includes a generator $D$ that can be thought of as a dilation. We find that this generator corresponds, in the bulk, to motion in the `holographic' bulk radial direction.

In section \ref{sec:entropy} we show that the entropy of the dS black hole may be obtained by averaging the quantum mechanical partition functions over one of the background metric components. The averaging we perform is natural: it corresponds to `gauge-fixing' the bulk time-diffeomorphism invariance by fixing the trace of the extrinsic curvature of the Wheeler-DeWitt slices \cite{York:1972sj, Witten:2018lgb, Witten:2022xxp}. It may be that, when the dust settles, this averaged partition function will play a more fundamental role in dS holography than the individual Hamiltonians $H(h)$.

Our work below connects to several important contemporary and long-standing themes in de Sitter holography. One of these is the dichotomy between the `meta-observer' \cite{Hartle:1983ai, Strominger:2001pn, Maldacena:2002vr, Anninos:2011ui, Anninos:2017eib, Chakraborty:2023yed, Chakraborty:2023los} and the static patch \cite{Gibbons:1977mu, Goheer:2002vf, Banks:2003cg, Alishahiha:2004md, Parikh:2004wh, Banks:2006rx, Dong:2010pm, Susskind:2011ap, Castro:2011xb, Anninos:2011af, Dong:2018cuv, Anninos:2020hfj, Susskind:2021dfc}, and the need to add an observer with a `time' to the static patch \cite{Chandrasekaran:2022cip}. Our family of quantum theories on worldtubes may be closely related to ideas connecting families of partition functions and wavefunctions using `$T^2$' type deformations \cite{McGough:2016lol, Hartman:2018tkw, Coleman:2021nor, Araujo-Regado:2022gvw, Araujo-Regado:2022jpj}. Worldtubes in the static patch have previously been considered in \cite{Anninos:2011jp, Anninos:2011zn, Witten:2023qsv}.

Our construction is inspired by an analogous correspondence between the Wheeler-DeWitt states of black hole interiors and asymptotic AdS/CFT partition functions \cite{Hartnoll:2022snh}. In an AdS/CFT setting one wishes to understand the black hole interior, and in particular the strongly curved singularity, starting from the well-defined boundary theory. In the cosmological context we have argued in the opposite direction, from a weakly curved asymptotically dS cosmological interior out to the static patch. Setups in which a dS cosmological interior can be engineered with AdS asymptotics, such as \cite{Anninos:2017hhn}, potentially have the best of both worlds and may provide a fruitful laboratory to further extend the ideas that we develop below.

\section{dS-Schwarzschild from the Hamilton-Jacobi equation}

The classical limit of the Wheeler-DeWitt equation is the gravitational Hamilton-Jacobi equation. It will therefore be useful to first recover the classical dS-Schwarzschild black hole using the Hamilton-Jacobi formulation of gravity. Our presentation here will follow \cite{Hartnoll:2022snh} closely, with a few improvements.

The action for pure, four dimensional gravity with a positive cosmological constant is
\be\label{eq:Ifull}
I[g] = \frac{1}{16 \pi G_N} \int d^4x \sqrt{-g} \left( R - \frac{6}{L^2} \right) + \frac{1}{8 \pi G_N} \int d^3x \sqrt{h} K \,.
\ee
Here $G_N$ is Newton's constant and $L$ is the de Sitter radius. The final term is the Gibbons-Hawking boundary term for the induced boundary metric $h$. Throughout this paper we will be concerned with the `minisuperspace' ansatz for the metric:
\be\label{eq:mini}
ds^2 = - N^2 dr^2 + \frac{g_{tt} dt^2}{(\Delta t)^2} + R^2 d\Omega^2_{S^2} \,.
\ee
Here $d\Omega^2_{S^2}$ is the metric on a two-sphere and $N$, $g_{tt}$ and $R$ are functions of $r$ only. We have explicitly included a factor of $\Delta t$, the extent of the $t$ coordinate, in the metric. This amounts to a pre-emptive rescaling of $g_{tt}$ that will simplify formulae in the remainder. The notation in (\ref{eq:mini}) indicates that this is a description of the `cosmological interior' in Fig.~\ref{fig:penrose}, where $r$ is the timelike coordinate. Evaluating the action on this ansatz gives
\be\label{eq:lag}
I[N,g_{tt},R] = \int \frac{dr {\mathcal L}}{2 G_N} \,,
\ee
where the Lagrangian density
\be
{\mathcal L} = 
N \sqrt{g_{tt}}\left(1 - \frac{3 R^2}{L^2} \right) - \frac{(\pa_r R) (\pa_r (R g_{tt}))}{N \sqrt{g_{tt}}} \,.
\ee
This minisuperspace theory has been shown to possess a Schr\"odinger symmetry \cite{BenAchour:2023dgj}.

From the action (\ref{eq:lag}) we obtain the conjugate momenta $\{\pi_{tt}, \pi_R\}$ and the Hamiltonian in the usual way. The Lagrange multiplier $N$ is then seen to impose the Hamiltonian constraint
\be\label{eq:ham}
(R \pi_R - g_{tt} \pi_{tt}) \pi_{tt} - \frac{R^2}{(2 G_N)^2} \left(\frac{3 R^2}{L^2} - 1\right) = 0 \,.
\ee
The Hamiltonian-Jacobi equation is obtained by introducing a function $S(g_{tt},R)$ such that $\pi_R = \pa_R S$ and $\pi_{tt} = \pa_{g_{tt}} S$. From (\ref{eq:ham}) we have
\be(R \pa_R S - g_{tt} \pa_{g_{tt}} S) \pa_{g_{tt}} S - \frac{R^2}{(2 G_N)^2} \left(\frac{3 R^2}{L^2} - 1\right) = 0 \,. \label{eq:HJ}
\ee
To obtain the classical solution to the equations of motion from the Hamilton-Jacobi equation one must, as usual, find a solution to (\ref{eq:HJ}) that has a nontrivial constant of integration. The overall shift in $S$ by a constant trivially drops out and will not be considered further. In fact, one can find two different solutions for $S$. These lead to the same classical solution to the equations of motion, and will later be related by a Fourier transform of the semiclassical wavefunction. The two solutions are
\be\label{eq:S1}
S_1(g_{tt},R; c) = \frac{1}{2 G_N} \left(\frac{g_{tt} R}{c} + c R \left(\frac{R^2}{L^2}-1\right) \right) \,,
\ee
where $c$ is the constant of integration, and
\be\label{eq:S2}
S_2(g_{tt},R; M) = \frac{R}{G_N} \sqrt{-g_{tt} f_M(R)} \,, \qquad f_M(R) \equiv 1 - G_N \frac{2 M}{R} - \frac{R^2}{L^2} \,. 
\ee
where $M$ is now the constant of integration. There is a freedom to choose the overall sign in $S$ in the expressions above.

The classical solution is then obtained by setting
\be\label{eq:class}
\pa_c S_1 = - M \qquad \text{or} \qquad \pa_M S_2 = c\,.
\ee
Either of these equations leads to (for the latter equation this requires $c>0$ when $g_{tt} > 0$, and indeed we want $c>0$ for solutions in the future cosmological interior)
\be\label{eq:gttsol}
g_{tt} = - c^2 f_M(R) \,.
\ee
The equation of motion for $N$ following from the action (\ref{eq:lag}), using (\ref{eq:gttsol}) to eliminate $g_{tt}$, gives
\be\label{eq:Nsol}
N^2 dr^2 = -\frac{dR^2}{f_M(R)} \,.
\ee
Putting (\ref{eq:gttsol}) and (\ref{eq:Nsol}) into
the ansatz (\ref{eq:mini}) gives the dS-Schwarzschild spacetime in familiar coordinates
\be\label{eq:sol}
ds^2 = - f_M(R) d \left(\frac{c t}{\Delta t}\right)^2 + \frac{dR^2}{f_M(R)} + R^2 d\Omega^2_{S^2} \,.
\ee
The constant $M$ is recognized as the mass of the black hole while $c$ determines the normalization of the `time' coordinate $t$. We may recall that in the cosmological interior $f_M(R) < 0$ so that $R$ is timelike.

We emphasize that (\ref{eq:gttsol}) is a `timeless', relational way to express the dS-Schwarzschild spacetime. It gives one metric component directly in terms of another, with no reference to coordinates on the spacetime.

\section{Semiclasscal Wheeler-DeWitt states}

The classical discussion of the previous section is the starting point for a description of semiclassical quantum states. The Wheeler-DeWitt equation \cite{DeWitt:1967yk} is a canonical quantization of the Hamiltonian constraint (\ref{eq:ham}). The momenta are promoted to operators obeying the usual commutation relations, so that the wavefunction $\Psi(g_{tt},R)$ obeys
\be\label{eq:wdw}
- \frac{\pa}{\pa g_{tt}} \left(g_{tt} \frac{\pa \Psi}{\pa g_{tt}} - R \frac{\pa \Psi}{\pa R}\right) + \frac{R^2}{(2 G_N)^2} \left(\frac{3 R^2}{L^2} - 1\right) \Psi = 0 \,.
\ee
There is an ordering ambiguity in the quantization of (\ref{eq:ham}). This will not be important for the leading order semiclassical physics that we will be concerned with. We have fixed the ordering in (\ref{eq:wdw}) by identifying the differential operator in the Wheeler-DeWitt equation with the Laplacian on the inverse DeWitt metric \cite{Halliwell:1988wc, Halliwell:1989myn}.

It is clear that semiclassical solutions to (\ref{eq:wdw}) with $\Psi = e^{iS}$ will be given to leading order by solutions to the Hamilton-Jacobi equation (\ref{eq:HJ}). The Wheeler-DeWitt equation should not be trusted beyond the semiclassical regime, as pure gravity is not a microscopically complete theory. Nonetheless, it is straightforward to solve (\ref{eq:wdw}) exactly so we will do this for completeness. We will then take the semiclassical limit of the exact solutions. Mirroring our classical discussion in the previous section, there are two natural ways to write the general exact solution to (\ref{eq:wdw}). Firstly one may write
\be\label{eq:Psi1}
\Psi(g_{tt},R) = \int_{-\infty}^\infty \frac{dc}{2\pi} \beta(c) e^{i S_1(g_{tt},R;c)} \,.
\ee
Here $\beta(c)$ is an arbitrary function and $S_1$ is the Hamilton-Jacobi function given above in (\ref{eq:S1}). The semiclassical solution turns out to be exact in this case.
Allowing $c$ to be negative in (\ref{eq:Psi1}) means that we do not need to separately consider the $e^{\pm i S}$ solutions. That is to say, (\ref{eq:Psi1}) is the general solution and
\be\label{eq:basis}
\Psi(g_{tt},R;c) \equiv e^{i S_1(g_{tt},R;c)} \,,
\ee
is a basis of solutions. As we have said above, we will be considering solutions with $c>0$ that correspond to the future cosmological interior.

Secondly, one may Fourier transform (\ref{eq:Psi1}) by setting
\be
\beta(c) = \int \frac{dM}{\sqrt{2\pi}} \alpha(M) e^{i M c} \,.
\ee
Here $\alpha(M)$ is a function and we can be agnostic at this point about the range of $M$ in the integral. We will want regular semiclassical states to be localized at non-negative $M$ that is furthermore below the Nariai limit. The solution (\ref{eq:Psi1}) then becomes
\be\label{eq:exact}
\Psi(g_{tt},R) = - g_{tt} \sqrt{2\pi} \int \frac{dM}{2\pi} \alpha(M) \frac{J_1\left({\textstyle \frac{R}{G_N}} \sqrt{- g_{tt} f_M(R)} \right)}{\sqrt{- g_{tt} f_M(R)}} \,.
\ee
Here $J_1$ is a Bessel function. In the rapidly oscillating semiclassical limit (\ref{eq:exact}) becomes
\be\label{eq:semi1}
\Psi(g_{tt},R) \to \frac{2 g_{tt}}{\sqrt{R}} \int \frac{dM}{2\pi} \frac{\sqrt{G_N} \, \alpha(M)}{(-g_{tt} f_M(R))^{3/4}} \left[ e^{i S_2(g_{tt},R;M)+ i \pi/4} + e^{-i S_2(g_{tt},R;M) - i \pi/4}\right] \,.
\ee
Here $S_2$ is the second Hamilton-Jacobi function that we found previously in (\ref{eq:S2}). The two terms in (\ref{eq:semi1}) originate from negative and positive $c$, respectively. They therefore separately solve the Wheeler-DeWitt equation. It is helpful to introduce the semiclassical solutions
\be\label{eq:alpha}
\Psi_\pm(g_{tt},R) = \int \frac{dM}{2\pi} \alpha(M) e^{\pm i S_2(g_{tt},R;M)} \,.
\ee
The upshot is that one may write the general semiclassical solution as a superposition of either $e^{i S_1(t_{tt},R;c)}$ states or of $e^{i S_2(t_{tt},R;M)}$ states. We will only be working to leading order in the semiclassical limit in the remainder, and therefore we have dropped all prefactors of the exponential terms in (\ref{eq:alpha}).
The split into positive and negative modes can also be done at the level of the exact Bessel functions in (\ref{eq:exact}), with $J_1$ written as a sum of two $K_1$ Bessel functions. Solutions with $e^{\pm i S_2}$ correspond to growing and collapsing universes, differing in the sign of $\pi_{tt}$ and $\pi_R$. For some purposes it is natural to consider superpositions of expanding and collapsing universes, but due to decoherence they may effectively be considered separately, e.g.~\cite{Banks:1984cw, Halliwell:1989vw, Padmanabhan:1989rm}. More detailed discussion of related issues in a similar context may be found in \cite{Hartnoll:2022snh}.

As is familiar from standard WKB wavefunctions in quantum mechanics, classical solutions are recovered by building wavepackets. For example if we take the Gaussian wavepacket
\be\label{eq:beta}
\beta(c) = \exp\left\{i M_o c - \frac{\Delta^2}{2} (c-c_o)^2 \right\} \,,
\ee
then, performing the $c$ integral by stationary phase, the $\Psi$ wavefunction in (\ref{eq:Psi1}) will be strongly peaked on $g_{tt}$ and $R$ such that
\be
 \left. \frac{\pa S_1}{\pa c} \right|_{c = c_o} = - M_o \,.
\ee
Which is to say that the wavefunction is strongly supported on the classical solution (\ref{eq:class}) with $M=M_o$ and $c = c_o$. More explicitly, to leading semiclassical order the wavefunction (\ref{eq:Psi1}) becomes
\be\label{eq:onshell}
\Psi_{c_o,M_o}(g_{tt},R) = \delta\left(g_{tt} + c_o^2 f_{M_o}(R) \right) \exp\left\{- \frac{i c_o}{G_N} R f_{M_o}(R)\right\} \,.
\ee
Here we have taken the limit in which the strongly peaked Gaussian that arises after performing the $c$ integration becomes a delta function.
The phase in (\ref{eq:onshell}) is equal to the on-shell action of dS-Schwarzschild space integrated up to $R$ (up to a constant of integration that depends on the other endpoint of the integral). Solutions to the Hamilton-Jacobi equation give the on-shell action on general grounds, so this is a check of our algebra.

It will be important to appreciate that in the quantum theory the two classical constants of integration $\{c,M\}$ arise in different ways. One of the constants parameterises a basis of quantum states. The other constant arises in the stationary phase approximation to wavepackets, which recovers the classical phase space. By choosing different bases of states, the roles of the two constants can be exchanged.

\section{Static patch partition functions}
\label{sec:static}

At the cosmological horizon $g_{tt} \to 0$ and the minisuperspace metric (\ref{eq:mini}) becomes degenerate. This is well-known to be a coordinate artifact. The coordinates used in the metric (\ref{eq:mini}) are only able to describe slices of spacetime that are localized in the cosmological interior. There certainly exist spacelike slices of dS-Schwarzschild that cut through the cosmological horizon, but these should be described by different coordinates. Such slices will necessarily break the symmetry of dS-Schwarzschild given by shifts in $t$. We wish to retain this symmetry, that underpins the minisuperspace description, and hence we will do something different.

It is also well-known that the coordinate $t$ can be extended through horizons at the cost of it acquiring a constant imaginary shift, e.g.~\cite{Fidkowski:2003nf, Hartman:2013qma}. Within our time-independent minisuperspace, at least, this shift does not seem to be important. For example, at a classical level the relational form of the dS-Schwarzschild given by (\ref{eq:gttsol}) continues without change from $g_{tt} > 0$ in the cosmological interior to $g_{tt} < 0$ in the `static patch'. A point made in \cite{Hartnoll:2022snh} is that continuation in $g_{tt}$ appears to be more straightforward than continuation in $t$ itself. In particular, the Hamilton-Jacobi function (\ref{eq:S1}) is linear in $g_{tt}$. For this reason, we will employ the Wheeler-DeWitt state in the form (\ref{eq:Psi1}), with the exponent given by (\ref{eq:S1}).

In the `static patch', where $g_{tt} < 0$, the fixed $r$ slices of the minisuperspace metric (\ref{eq:mini}) are Lorentzian. See Fig.~\ref{fig:penrose} above. In this region the wavefunction is more naturally interpreted as a Lorentzian partition function of a `boundary' theory, in the spirit of the AdS/CFT correspondence \cite{Gubser:1998bc, Witten:1998qj}. Partition functions are traces over a Hilbert space, which is quite distinct from a wavefunction. In the remainder of this section we give a proposal for how, with $g_{tt} <0$, $\Psi$ can be interpreted as a partition function on the boundary worldtube. Because there is no rigid boundary in the static patch, one expects to find a family of partition functions for the different possible geometries of the boundary worldtube.

We would first like to interpret the constant $M$, the mass of the dS-Schwarzschild black hole, as the expectation value of a dual quantum mechanical `Hamiltonian' $H(g_{tt},R)$:
\be\label{eq:Hvev}
\langle H \rangle = M = - \frac{\pa S_1}{\pa c} \,.
\ee
We may use the explicit expression (\ref{eq:S1}) for $S_1$ to obtain
\be\label{eq:Hvev2}
\langle H \rangle = \frac{g_{tt} R}{2 G_N}\frac{1}{c^2} + \frac{R}{2 G_N} \left(1 - \frac{R^2}{L^2} \right) \,.
\ee
Furthermore using (\ref{eq:S1}) to obtain
the conjugate momentum $\pi_{tt} = \pa_{g_{tt}} S_1$ we may verify that
\be\label{eq:S1again}
S_1 - 2 g_{tt} \pi_{tt}  = - c \langle H \rangle \,.
\ee
This is a useful expression because it will allow us to relate the semiclassical wavefunction $e^{i S_1}$ to a partition function where $H$ generates translations in $c$. We will see in (\ref{eq:full}) below that the extra term $2 g_{tt} \pi_{tt}$ in (\ref{eq:S1again}) is a boundary term in a gravitational path integral. Similar additional terms, relating wavefunctions and partition functions, have appeared in discussions of $T^2$ deformations \cite{Araujo-Regado:2022gvw, Araujo-Regado:2022jpj}.

We propose that the shifted Hamilton-Jacobi function in (\ref{eq:S1again}) may now be understood as the semiclassical limit of a quantum mechanical partition function $Z$ by upgrading\be\label{eq:corr}
e^{i [S_1(g_{tt},R; c)- 2 g_{tt} \pi_{tt}(g_{tt},R; c)]} \to \Tr \left[e^{- i c H(g_{tt},R)}\right] \equiv  Z_\text{QM}(g_{tt},R; c)\,.
\ee
The quantum mechanical trace in (\ref{eq:corr}) is understood to be normalized, with $\Tr \mathbb{1} = 1$.
In the classical limit, then, $\Tr \left[e^{- i c H}\right] = e^{- i c \langle H \rangle}$, recovering (\ref{eq:S1again}).
It is important to appreciate here that in the quantum mechanical description $H$ is an operator that depends on background fields $g_{tt}$ and $R$ but does not depend explicitly on `time' $c$. Instead, in the quantum mechanics, $H$ generates translations in $c$. Its expectation value must therefore be independent of $c$. The gravitational expression (\ref{eq:Hvev2}) for $\langle H \rangle$ does depend on $c$ explicitly. We know from the Hamilton-Jacobi relation (\ref{eq:Hvev}), however, that on classical solutions $\langle H \rangle$ is a constant independent of $c$. This fact is essential for the proposed correspondence to make sense.

Considering $e^{i S_1}$ as the semiclassical limit of a gravitational wavefunction $\Psi$, the proposed quantum duality (\ref{eq:corr}) can be rewritten explicitly as
\be\label{eq:duality}
\Psi(g_{tt},R;c)e^{-i 2 g_{tt} \langle\pi_{tt}\rangle_\Psi} = Z_\text{QM}(g_{tt},R; c) \,.
\ee
The left hand side involves a basis of quantum gravitational minisuperspace states of the cosmological interior, solutions to the WDW equation labelled by the constant of integration $c$. The right hand side is a quantum mechanical partition function with time $c$. The quantum mechanical theory lives on a world-tube with background metric labelled by $\{g_{tt},R\}$. The different possible backgrounds give a family of partition functions.

It should be emphasized that (\ref{eq:duality}) is not linear in $\Psi$. This means it is not immediately obvious how to deal with general wavefunctions of the form (\ref{eq:Psi1}), where the $\Psi(g_{tt},R;c)$ are linearly weighted by an arbitrary function $\beta(c)$. Our proposal is that (\ref{eq:duality}) holds for general gravitational states $\Psi$, so that:
\be\label{eq:Zgen}
\Psi(g_{tt},R)e^{-i 2 g_{tt} \langle\pi_{tt}\rangle_\Psi} = Z_\text{QM}(g_{tt},R) \equiv \int \frac{dc}{2 \pi} \beta(c) Z_\text{QM}(g_{tt},R; c) \,.
\ee
Here $\beta(c)$ is defined via the decomposition (\ref{eq:Psi1}) of $\Psi$ in the $\Psi(g_{tt},R;c)$ basis.
The more general form (\ref{eq:Zgen}) of the correspondence can be motivated by expressing the expectation value in the exponent on the left hand side of (\ref{eq:Zgen}) as a boundary term in a bulk partition function. We may write the correspondence (\ref{eq:Zgen}) as an equivalence of partition functions:
\be\label{eq:path}
Z_\text{QM}(g_{tt},R) = Z_\text{grav}(g_{tt},R) \,,
\ee
where
\be\label{eq:full}
Z_\text{grav}(g_{tt},R) \equiv \left.  \int {\mathcal D}g \exp\left\{i \left[\left(\int^{r_\star} \frac{dr {\mathcal L}}{2 G_N} \right) - 2 g_{tt}(r_\star) \pi_{tt}(r_\star) \right]\right\} \right|^{g(r_\star) = \{g_{tt},R\}}_\beta \,.
\ee
In the gravitational path integral $\pi_{tt} = \frac{1}{2 G_N} \frac{\pa {\mathcal L}}{\pa (\pa_{r}g_{tt})}$. The 
subscript $\beta$ instructs us to impose boundary conditions away from $r=r_\star$ that, in the absence of the $g_{tt} \pi_{tt}$ term, would prepare the solution $\Psi(g_{tt},R)$ to the Wheeler-DeWitt equation. The equivalence of partition functions in (\ref{eq:path}) has the same form as the AdS/CFT dictionary \cite{Gubser:1998bc,Witten:1998qj}.

As an application and also a check of (\ref{eq:Zgen}), consider the Gaussian wavepacket state $\Psi_{c_o,M_o}(g_{tt},R)$ with $\beta(c)$ given in (\ref{eq:beta}). We wrote this state explicitly in (\ref{eq:onshell}). According to (\ref{eq:Zgen}), this state corresponds to the partition function
\be\label{eq:Znice}
Z_{c_o,M_o}(g_{tt},R) = \Tr \left[\delta(H - M_o)\right] \,.
\ee
The right hand side of (\ref{eq:Znice}) is independent of $c_o$; we explain how to deal with the $c_o$ dependence of the left hand side shortly. As in (\ref{eq:onshell}) we have taken the delta function limit of the Gaussian that arises after performing the integral over $c$ in (\ref{eq:Zgen}). We are working to leading semiclassical order and therefore not keeping track of overall prefactors. It is clear that (\ref{eq:Znice}) describes a microcanonical partition function, where the energy $H$ has been fixed to $M_o$.  At the leading semi-classical level the expression (\ref{eq:Znice}) for the partition function must equal the continuation of the wavefunction (\ref{eq:onshell}) to $g_{tt} < 0$. We now proceed to verify that this is the case. Noting that on the classical solution
\be\label{eq:gttptt}
2 g_{tt} \langle \pi_{tt} \rangle = - \frac{c_o}{G_N} R f_{M_o}(R) \,,
\ee
we verify that the exponent in the correspondence (\ref{eq:Zgen}) precisely cancels the phase in (\ref{eq:onshell}). It follows from (\ref{eq:Zgen}), (\ref{eq:onshell}) and (\ref{eq:Znice}) that we must have
\be\label{eq:deltadelta}
\Tr \Big[\delta(H(g_{tt},R) - M_o)\Big] =\delta\Big(g_{tt} + c_o^2 f_{M_o}(R) \Big) = \delta\Big(g_{tt} + \frac{R^2 f_{M_o}(R)}{4 G_N^2 \langle \pi_{tt}\rangle_{M_o}^2} \Big) \,.
\ee
In the final step we used (\ref{eq:gttptt}) to re-express $c_o$ in terms of the expectation value (\ref{eq:gttptt}) of $\pi_{tt}$. This can be though of as the expectation value of an operator $\pi_{tt}$ in the state of the quantum mechanics with energy $M_o$. Once again, we are not keeping track of prefactors.
The equality (\ref{eq:deltadelta}) is verified using the classical expression for $\langle H \rangle$ in (\ref{eq:Hvev2}). From the perspective of our duality, (\ref{eq:deltadelta})
is the statement that there is a state in the quantum mechanical theory with energy $M_o$ if and only if $g_{tt}$ and $R$ are related by the corresponding classical gravitational equations of motion with constant of integration $M_o$. Equation (\ref{eq:deltadelta}) expresses an essence of the correspondence: a quantum mechanical Hamiltonian $H$ with the property (\ref{eq:deltadelta}) is equivalent to a semiclassical dS-Schwarzschild universe. This demonstrates the possibility of associating a (non-vanishing) Hamiltonian to the dynamics of closed universes.

From (\ref{eq:deltadelta}), we see that the spectrum of $H$ is bounded above. For $g_{tt} < 0$, the allowed values of $M_o$ obey\be
\label{eq:massbound}
M_o = \frac{R}{2 G_N} \left(1 - \frac{R^2}{L^2} \right) + \frac{2 G_N}{R} g_{tt} \langle \pi_{tt}\rangle_{M_o}^2 \leq \frac{R}{2 G_N} \left(1 - \frac{R^2}{L^2} \right) \leq \frac{1}{3 \sqrt{3}} \frac{L}{G_N}  \,.
\ee
This is, of course, just the familiar Nariai bound on the black hole mass.

\section{Symmetry algebra}
\label{sec:algebra}

In the dual quantum mechanical theory $g_{tt}$ and $R$ are background functions. For a given background, however, the gravitational momenta $\pi_{tt}$ and $\pi_R$ depend on time and hence should be upgraded to operators in the quantum mechanics. Their expectation values, computed using the classical gravitational theory, obey
\begin{align}
g_{tt} \frac{\pa \langle \pi_{tt} \rangle}{\pa c} & = - g_{tt} \frac{\pa \langle H \rangle}{\pa g_{tt}} = - \langle H \rangle + \frac{R}{2 G_N} \left(1 - \frac{R^2}{L^2} \right) \,, \label{eq:aa} \\
R \frac{\pa \langle \pi_{R} \rangle}{\pa c} & = - R \frac{\pa \langle H \rangle}{\pa R} = - \langle H \rangle + \frac{R}{G_N} \frac{R^2}{L^2} \,. \label{eq:bb}
\end{align}
The first equality in each line follows from the definitions $\langle H \rangle = - \pa S_1/\pa c$, $\langle \pi_{tt} \rangle = \pa S_1/\pa g_{tt}$, $\langle \pi_R \rangle = \pa S_1/\pa R$ and commuting derivatives. These first equalities demonstrate that the bulk gravitational degrees of freedom obey Hamilton's equations with $c$ as time and $\langle H \rangle$ as the classical Hamiltonian. The second equalities follow from, for example, the expression (\ref{eq:Hvev2}) for $\langle H \rangle$.
Equations (\ref{eq:aa}) and (\ref{eq:bb}) show that the $c$ derivatives of $\pi_{tt}$ and $\pi_R$ are independent of $c$ on classical solutions. This is important for what we are about to do next.

Given that $H$ generates translations in time $c$ in the quantum mechanical theory, the relations (\ref{eq:aa}) and (\ref{eq:bb}) require (semiclassically) the quantum mechanical commutators
\begin{align}
i[H,g_{tt} \pi_{tt}] & = - H + \frac{R}{2 G_N} \left(1 - \frac{R^2}{L^2} \right) \,, \label{eq:one}\\
i[H,R \pi_R] & = - H + \frac{R}{G_N} \frac{R^2}{L^2} \,. \label{eq:two}
\end{align}
The remaining commutator between the three operators $H, \pi_{tt}$ and $\pi_R$ can be constrained using the Jacobi identity
\begin{align}
i [H,i [g_{tt} \pi_{tt},R \pi_{R}]] & = - i [g_{tt} \pi_{tt}, i [R \pi_{R},H]] - i [R \pi_{R}, i [H,g_{tt} \pi_{tt}]] \\
& = \frac{R}{2 G_N} \left(1 - \frac{3 R^2}{L^2} \right) \,.
\end{align}
The simplest possibility for the commutator, involving only the operators we have at our disposal at this point, is then
\be
i [g_{tt} \pi_{tt},R \pi_{R}] = g_{tt} \pi_{tt} - R \pi_R + \a H + \b \,. \label{eq:three}
\ee
Here $\a$ and $\b$ are undetermined c-numbers. These undetermined terms can be removed by letting $g_{tt} \pi_{tt} \to g_{tt} \pi_{tt} + \gamma + \delta H$, with suitably chosen constants $\gamma, \delta$, without changing either of the other commutators (\ref{eq:one}) and (\ref{eq:two}). Therefore we may set $\a = \beta = 0$ in (\ref{eq:three}).

The commutators (\ref{eq:one}), (\ref{eq:two}) and (\ref{eq:three}) generate an interesting algebra. Let
\be
P \equiv H - \frac{R}{G_N} \frac{R^2}{L^2} \,, \qquad
X \equiv R \pi_R - g_{tt} \pi_{tt} \,, \qquad
D \equiv R \pi_R \,.\label{eq:RdR}
\ee
Then the algebra becomes
\be\label{eq:algebra}
i [X,D] = X \,, \qquad i [P,D] = - P \,, \qquad i [X,P] = h(R) \,,
\ee
which may be recognized as a central extension of the 1+1 dimensional Poincar\'e algebra $\mathfrak{e}(1,1)$. In $\mathfrak{e}(1,1)$, without the extension, $D$ would generate Lorentz boosts, while $X \pm P$ would generate spacetime translations. The coefficient in the central extension is
\be\label{eq:coefficient}
h(R) = \frac{R}{2 G_N} \left(1 - \frac{3 R^2}{L^2} \right) \,.
\ee
This coefficient is positive for $R < \frac{L}{\sqrt{3}}$, which always contains the black hole horizon, and negative for $R > \frac{L}{\sqrt{3}}$, which always contains the cosmological horizon. The two horizons coincide in the Nariai limit at $R = \frac{L}{\sqrt{3}}$, where the coefficient (\ref{eq:coefficient}) vanishes.

The representation of the algebra in which the quantum mechanical states transform is constrained by the invariant \cite{Patera:1976ud}
\be\label{eq:inv}
C \equiv \frac{PX + XP}{2 h(R)} - D \,.
\ee
Using the classical gravitational values for the momenta and energy we obtain
\be\label{eq:CC}
\langle C \rangle = \frac{\langle X \rangle \langle P \rangle}{h(R)} - \langle D \rangle = - 2 g_{tt} \langle \pi_{tt} \rangle \,.
\ee
Curiously, the invariant thereby gives a group-theoretic interpretation to the phase of the gravitational wavefunction (\ref{eq:onshell}), via (\ref{eq:gttptt}). Recall that this phase was cancelled out of the quantum mechanical partition function in the correspondence (\ref{eq:Zgen}).

The algebra (\ref{eq:algebra}) has been studied previously following its introduction as the gauge symmetry of a 1+1 dimensional theory of gravity by Cangemi and Jackiv \cite{Cangemi:1992bj}. Aspects of its representation theory have been discussed in \cite{negro1993local, Cangemi:1993sd, deMello:2002zkl, Popovych:2003xb, biggs}. An explicit field theory representation of the algebra was given in \cite{Karat:1998qg}, as the symmetry group of a complex scalar field in 1+1 dimensions in a constant background electric field. A simple realization of the algebra (\ref{eq:algebra}) on functions of a single variable $x$ is
\be\label{eq:simple}
X = x \,, \qquad P = h(R) p \,, \qquad D = \half \left(x p + p x \right) \,.
\ee
Here $p = - i \pa_x$ is the usual momentum operator.
In this realization the generator $D$ is manifestly a dilation. The correspondence between dilation in quantum mechanics and motion in the bulk `radial' direction, implied by (\ref{eq:RdR}), has a holographic flavor.
The simple realization (\ref{eq:simple}) is not the appropriate one for our dual theory because the invariant (\ref{eq:inv}) vanishes. However, (\ref{eq:simple}) can be extended to obtain a non-vanishing invariant by adding an operator to $D$ that commutes with both $P$ and $X$.

More information from the gravitational theory, beyond superspace, is needed to identify the appropriate representation of the algebra for our dS-Schwarzschild context. In addition to (\ref{eq:CC}), a further constraint follows from the Nariai bound (\ref{eq:massbound}) on the black hole mass. This requires states dual to classical gravity solutions to have
\be
- \frac{R}{G_N} \frac{R^2}{L^2} \leq \langle P \rangle \leq h(R) \,. 
\ee
The lower bound is from positivity of the black hole mass.

Two copies of the extended Poincar\'e algebra appears as subalgebras of the Schr\"odinger symmetry discussed by \cite{BenAchour:2023dgj}. We are unsure at this point whether there is a connection between that symmetry and ours, which act on different Hilbert spaces.

\section{The black hole entropy from averaging}
\label{sec:entropy}

One might expect that if the boundary worldtube is taken close to the black hole horizon, then the quantum mechanical theory should know about the black hole microstates. In this section we show that this is indeed the case.

As written, the microcanonical trace (\ref{eq:deltadelta}) does not seem to know about black hole entropy. One would like to see a factor of $e^{S_\mathcal{H}}$, with $S_\mathcal{H}$ the black hole entropy, on the right hand side of (\ref{eq:deltadelta}). It turns out that an averaging over the quantum mechanical theory is needed. To see how this works, let us first extract the black hole entropy from the gravitational on-shell action.
We will do this in the spirit of \cite{Banados:1993qp,Teitelboim:2001skl}, in which
the on-shell action is supplemented by an additional boundary term at the horizon. 

The delta function in the Wheeler-DeWitt state (\ref{eq:onshell}), that subsequently appears on the right hand side of (\ref{eq:deltadelta}), suggests that it is natural to integrate the state against one of the metric functions. A specific form of this integral can be motivated as follows. The Wheeler-DeWitt equation itself represents a gravitational gauge redundancy. As has been emphasized recently \cite{Witten:2018lgb, Witten:2022xxp}, and known for a long time \cite{York:1972sj}, a natural way to fix this redundancy is to consider slices with fixed trace $K$ of the extrinsic curvature. The trace $K \propto \pi_v$ is conjugate to the local volume
\be
v \equiv \sqrt{- g_{tt}} R^2 \,,
\ee
of the induced metric (with $g_{tt}<0$). To express the wavefunction in terms of extrinsic curvatures rather than metric components, one must Fourier transform the wavefunction from $\Psi(v,[h])$ to $\widetilde \Psi(\pi_v,[h])$. Here $[h] \equiv g_{tt}/R^2$ is the conformal class of the induced metric. To fix the trace of the extrinsic curvature, one can then evaluate the Fourier transformed wavefunction on the expectation value in the original state, i.e.~let $\widetilde \Psi([h]) \equiv \widetilde \Psi(\langle \pi_v \rangle_\Psi,[h])$. In this way we obtain a wavefunction $\widetilde \Psi([h])$ of the conformal data only.

To implement the Fourier transform from $v$ to $\pi_v$ we need to add a term proportional to $i v \pi_v$ in the exponent. A natural term to add is the Gibbons-Hawking boundary term:
\be
\frac{1}{8 \pi G_N} \int d^3x \sqrt{h} K = 
g_{tt} \pi_{tt} + \frac{1}{2} R \pi_R =
\frac{3}{2} v \pi_v  \,.
\ee
In fact we already included this term in our starting point (\ref{eq:Ifull}), so what we will do now is to subtract it off. That is, we consider the following partial integration of the gravitational wavefunction $\Psi(g_{tt},R)$:
\be\label{eq:transform}
\widetilde \psi(R) \equiv \int dg_{tt} \, e^{- i \frac{3}{2} v \langle \pi_v \rangle_\Psi} \Psi(g_{tt},R) = \int dg_{tt} \, e^{- i (g_{tt} \langle \pi_{tt} \rangle_\Psi + \frac{1}{2} R \langle \pi_R \rangle_\Psi)} \Psi(g_{tt},R) \,.
\ee
The integral in (\ref{eq:transform}) is precisely the Fourier transform discussed in the previous paragraph, evaluated on the original expectation value $\langle \pi_v \rangle_\Psi$ as we described, with the difference that we have integrated over $g_{tt}$ rather than $v$. The last fact means that the reduced wavefunction $\widetilde \Psi$ is a function of $R$ rather than the conformal class $[h]$. We do this because $R$ is single-valued on the static patch, increasing monotonically from the black hole horizon to the cosmological horizon, and is therefore a convenient coordinate. In contrast, $g_{tt}$ and the conformal data $[h] = g_{tt}/R^2$ are not single-valued in the static patch as they vanish at both horizons.

For the Gaussian wavepacket state (\ref{eq:onshell}) we obtain from (\ref{eq:transform}) that
\begin{align}
\widetilde \Psi_{c_o,M_o}(R) & = \int d g_{tt} \delta\left(g_{tt} + c_o^2 f_{M_o}(R) \right) \exp\left\{i c_o \frac{4 \pi R^2}{4 G_N} \frac{f'_{M_o}(R)}{4 \pi}\right\} \,, \label{eq:ww}\\
& = \exp\left\{\frac{i c_o}{\beta_R} \frac{A_R}{4 G_N}\right\} \,.
\end{align}
Here we defined
\be
A_R \equiv 4 \pi R^2 \,, \qquad \beta_R \equiv \frac{4 \pi}{f'_{M_o}(R)} \,.
\ee
If $R = R_\mathcal{H}$, the black hole horizon radius, then $\beta_{R_{\mathcal{H}}} = \beta_\mathcal{H}$ is the inverse Hawking temperature and $\frac{A_{R_{\mathcal{H}}}}{4 G_N} = S_\mathcal{H}$ is the Bekenstein-Hawking entropy. Thus, setting $c_o = -i \beta_\mathcal{H}$, we recover the black hole entropy as:
\be\label{eq:entropy}
\widetilde \Psi_{- i \beta_\mathcal{H},M_o}(R_\mathcal{H}) = e^{S_\mathcal{H}} \,.
\ee
In this expression $\beta_\mathcal{H}, R_\mathcal{H}$ and $S_\mathcal{H}$ all depend on $M_o$. The imaginary value of $c_o$ here is such that the resulting classical Euclidean solution (\ref{eq:sol}) has no conical defect at the black hole horizon. In Euclidean dS-Schwarzschild there must be a conical defect at either the cosmological or the black hole horizon. 
The steps we have just gone through show that to extract the black hole entropy from the transformed wavefunction at $R = R_\mathcal{H}$, the black hole radius, one should require regularity at the black hole horizon. One may analogously extract the entropy associated to the cosmological horizon from the transformed wavefunction at the cosmological horizon radius.

Equation (\ref{eq:entropy}) shows that the partially transformed wavefunction $\widetilde \Psi$ knows about black hole entropy in a natural way. We may now determine the corresponding quantity in the dual quantum mechanics.
Using the identification of delta functions (\ref{eq:deltadelta}) in the transformed state (\ref{eq:ww}) gives
\be\label{eq:obj}
\widetilde \Psi_{c_o,M_o}(R) = \int d g_{tt} e^{i [g_{tt} \langle\pi_{tt}\rangle_{c_o} - \frac{1}{2} R \langle\pi_{R} \rangle_{c_o}]} \Tr \left[\delta(H(g_{tt},R) - M_o)\right] \,.
\ee
The exponent comes from the transform (\ref{eq:transform}) together with the shift by $2g_{tt} \langle \pi_{tt} \rangle$ in (\ref{eq:corr}). The $c_o$ subscript tells us that these are expectation values of Heisenberg picture operators, evolved in $c$ by $H$. At leading semiclassical order these expectation values equal the values of the momenta on the classical gravitational solution. Setting $\langle \pi_{tt}\rangle = \pa S_1/\pa g_{tt}$, and similarly for $\langle \pi_R\rangle$, it is verified that the exponent then matches that in (\ref{eq:ww}).

It is remarkable that the transformed `gauge-fixed' gravitational wavefunction (\ref{eq:obj}) equals an average over dual quantum mechanical (microcanonical) partition functions. The averaging is over $g_{tt}$. 
Given that the worldtubes are in the midst of a quantum fluctuating geometry,  the averaging is perhaps natural and, indeed, such averaged theories can be thought of as $0+1$ dimensional theories of quantum gravity, as recently discussed in \cite{Anninos:2021ydw}.

The expression (\ref{eq:entropy}) shows that the black hole entropy can be extracted from the averaged partition function in (\ref{eq:obj}) if it is evaluated at the black hole radius $R = R_\mathcal{H}$ and at the appropriate imaginary value of $c_o = - i \beta_\mathcal{H}$. The black hole radius is picked out as the minimum radius for which the {\it unaveraged} semiclassical partition function (\ref{eq:deltadelta}) has support on $g_{tt} < 0$. The value of the imaginary time is picked out, as we explained above, by the absence of a conical singularity at the black hole horizon in the Euclidean gravitational solution. A conical singularity at the horizon means that the action is not extremized there, see e.g.~\cite{Banados:1993qp}. A careful treatment of this point should show, then, that the quantum wavepacket fails to localize on the classical solution at $R = R_\mathcal{H}$ unless the correct value of $c_o$ is chosen. From this fact it should be possible to determine $\beta_\mathcal{H}$, in principle, given the dual quantum mechanical Hamiltonians $H(g_{tt},R)$.

We end with a comment on the continuation of 
gravitational wavefunctions with the imaginary value $c_o = - i \beta_\mathcal{H}$ to the cosmological interior. This continuation is unnatural given the conical singularity at the cosmological horizon.  Further to this point, the wavefunction $\Psi_{- i \beta_\mathcal{H}, M_o}(g_{tt},R)$ in (\ref{eq:onshell}) grows exponentially as $R \to \infty$ towards the late cosmological interior because $f_{M_o}(R) \to - \infty$ in this limit. The DeWitt norm --- which is more like a flux than a conventional norm, see \cite{Hartnoll:2022snh} for a recent discussion --- vanishes on these purely real states. The averaged wavefunction $\widetilde \Psi_{- i \beta_\mathcal{H},M_o}(R)$ instead decays exponentially towards large $R$. All told, the entropy seems to be most easily discussed within the `pure static patch' perspective on the correspondence that we noted in the introduction.

\section{Discussion}

We have proposed that an AdS/CFT-like duality is possible for dS-Schwarzschild spacetimes, where one has a family of dual quantum mechanical theories defined on static patch worldtubes. Time translation in these dual theories corresponds to moving among different basis states of the gravitational theory. We argued that it was natural to gauge-fix the trace of the extrinsic curvature of the worldtubes by performing a certain average over the worldtube theories. The upshot of this procedure was that the black hole entropy could be extracted from the averaged partition function.

An important test of our proposal will be whether it can be extended beyond minisuperspace, to allow for inhomogeneous perturbations of the geometry. This extension, if it is possible, may also shed further light on the nature of the dual quantum mechanics. The inhomogeneous modes will have corresponding momenta that will become operators in the quantum mechanical theory. One would like to understand whether these enlarge the extended Poincar\'e algebra that we have uncovered. A natural candidate for the expectation value of the Hamiltonian, that played a central role in our analysis, in this more general context would be the Brown-York energy \cite{Brown:1992br} of the (now inhomogeneous) worldtube.

Another ingredient that is currently missing from our proposed correspondence is the role of entanglement between the two antipodal static patches. This is similar to the thermofield double picture of black holes in AdS/CFT \cite{Maldacena:2001kr} and
has played a prominent role in recent works including \cite{Dong:2018cuv, Shaghoulian:2021cef, Lin:2022nss, Chandrasekaran:2022cip, Cotler:2023xku}.
As occurs in the case of AdS/CFT, it 
is possible that considering entangled copies of the dual static patch partition functions may give a deeper understanding of the sense in which these prepare quantum states of the cosmological interior.

Finally, as noted several times above, while we anchored our discussion of Wheeler-DeWitt states in the cosmological interior, this was not strictly necessary. One can also work directly with gravitational partition functions defined in the static patch via a gravitational path integral. A partition function formulation may also be applicable in spacetimes with zero cosmological constant. In particular, the $L \to \infty$ limit of all of the expressions in our paper appears to be sensible.
This may suggest an approach to flat space holography, with the cosmological horizon then arising as a $\frac{1}{L}$ deformation of the quantum mechanical theories dual to asymptotically flat spacetimes.


\section*{Acknowledgements}

It is a pleasure to acknowledge inspiring discussions with Dionysios Anninos, Diego Hofman, Albert Law and Ronak Soni. The work of S.A.H.~is partially supported by Simons Investigator award $\#$620869 and by STFC consolidated grant ST/T000694/1. 
M.J.B.~was supported by a Gates Cambridge Scholarship ($\#$OPP1144).

\providecommand{\href}[2]{#2}\begingroup\raggedright\endgroup

\end{document}